\documentclass[a4paper,12pt]{article}
\usepackage[cp1251]{inputenc}
\usepackage{epsfig}
\usepackage[english]{babel}
\usepackage{amssymb}

\begin{document}
\title{Properties of Fourier spectrum of the signal, generated at the accumulation point of period-tripling bifurcations}
\author{O.B. Isaeva}
\date{}
\maketitle\begin{center} \emph{Institute of Radio-Engineering and
Electronics of RAS, Saratov Branch, \\ Zelenaya 38, Saratov,
410019, Russia \\ E-mail: IsaevaOB@info.sgu.ru}\end{center}

\begin{abstract}
Universal regularities of the Fourier spectrum of signal,
generated by complex analytic map at the period-tripling
bifurcations accumulation point are considered. The difference
between intensities of the subharmonics at the values of frequency
corresponding to the neighbor hierarchical levels of the spectrum
is characterized by a constant $\gamma=21.9$ dB?, which is an
analogue of the known value $\gamma_F=13.4$ dB, intrinsic to the
Feigenbaum critical point. Data of the physical experiment,
directed to the observation of the spectrum at period-tripling
accumulation point, are represented.
\end{abstract}

One of the simplest objects, demonstrating non-trivial dynamics,
is the system with discrete time, defined by the quadratic map. In
particular, the period-doubling bifurcations cascade,
characterized by the universal Feigenbaum properties, occures in
this system~\cite{Kuznetsov}. By generalization to the case of
complex variable the quadratic map
\begin{equation}\label{logistic}
 z_{n + 1}=\lambda-z_{n}^{2}, \quad \lambda, z \in \mathbb{C}
\end{equation}
demonstrates the fractal object, known as Mandelbrot set, on the
parameter plane (see fig.~1)~\cite{Peitgen}. This object is an
aggregate of "leaves", inside which the periodic dynamics takes
place in a restricted domain of phase space, and fractal pattern
corresponding to the restricted chaotic dynamics (restriction
means, that orbits of the map doesn't escape to infinity). It is
possible to define the paths at the complex parameter plane, which
traverse the Mandelbrot set "leaves" according the special order.
Following these paths one can obtain different
period-multiplication bifurcation
cascades~\cite{Cvitanovic,Golberg}, for example period-tripling
cascade. In the work of Golberg-Sinai-Khanin~\cite{Golberg} the
critical point of the period-tripling bifurcations accumulation
has been considered. Let us denote it as GSK point:
\begin{equation}\label{4}
\lambda_{c}=0.0236411685377+0.7836606508052 i.
\end{equation}

If one separate real and imaginary parts in the 1D complex map,
then one obtain the equivalent description of the dynamics by the
2D map of the pair of real variables. Notice, that this 2D map is
of special form, because the corresponded functions must satisfy
to the Cauchy-Rieman conditions (analyticity conditions). As it is
shown in~\cite{Isaeva1,Peckham}, even the small non-analytic
perturbation in the map leads to the drastic distortion of the
dynamics, in particular, to the destruction of the universal
properties of the period-tripling and other period-multiplication
cascades. In view of this fact the problem of possibility of
realisation of the phenomena of complex analytic dynamics in the
physical systems get a fundamental significance (see, for
example~\cite{Beck}). In paper~\cite{Isaeva2} the problem of
realization of the Mandelbrot set on the parameter plane of the
coupled systems is discussed. The coupling must have a special
form and provide the special  symmetry of the system, necessary to
the analyticity conditions implementation. Based on this idea, in
the paper~\cite{Isaeva3} the electronic analog device, modelling
the dynamics of the coupled logistic maps is proposed. In this
system the first observation of the Mandelbrot set in physical
experiment is carried out.

With the help of this real experimental device one can observe the
bifurcations, intrinsic to the complex maps, and, in particular,
the period-tripling bifurcations cascade.

Studying the dynamical regimes in different systems, it is
convenient to use the spectral properties of the generated signal.
Thereupon, let us address to the problem of the Fourier spectrum
of the signal, generated by the complex quadratic map at the
critical point GSK of the accumulation of the period-tripling
bifurcations. From the data of the renormalization group
analysis~\cite{Cvitanovic,Golberg} one can conclude, that the
infinite number of unstable cycles of periods $N=3^k$, where
$k=1,2,3,...$, exist at the GSK point. With $N\rightarrow\infty$
these cycles approximate the critical attractor more and more
precisely. Direct and reverse discrete Fourier transformations can
be defined as follows:
\begin{equation}\label{fourier}
\begin{array}{c}
z_n=\sum \limits_{m=0}^{N-1} c_m \exp\left(\frac{2\pi i}{N}mn
\right),  \\

c_m=\frac{1}{N}\sum \limits_{m=0}^{N-1} z_n \exp\left(-\frac{2\pi
i}{N}mn\right),
\end{array}
\end{equation}
where $z_n$ is the sequence of the values of dynamical variable,
generated at the GSK critical point. The starting element of this
sequence is the extremum of the map, i.e. $z_0=0$. Value $f=m/N$
corresponds to the frequency of the component $c_m$. Let us
introduce  the designation for the squared amplitude of the $m$-th
component $S(f)=S(m/N)=|c_m|^2$.

At figure~2~(a) the Fourier spectrum of the signal, generated at
the GSK point is shown. It is represented as dependence of the
harmonic's intensities in dB units, i.e. $10 \lg S$, on the
frequency $f$. Notice that spectrum demonstrates fractal nature:
there are infinite number of peaks at the frequencies
corresponding to the periods of the existing in GSK point unstable
cycles. Structure of the spectrum between two neighbor peaks are
repeated. As it is visible at the figure, the difference of
intensities of harmonics at the frequencies $1/3^k$ and
$1/3^{k+1}$ is approximately $\gamma=21.9$ dB. This value is an
analog of the known constant $\gamma=13.4$ dB intrinsic to the
Feigenbaum critical point~\cite{Kuznetsov}.

At the figure~2~(b) the same spectrum of the map~(1) at the  GSK
point, but in double logarithmic scale is shown. Moreover, the
spectral intensities are rationed to the corresponded frequencies
in power $\kappa=6.15$ (this constant is selected empirically).
With such representation the self-similar structure of the
spectrum looks more clear. It is easy to see, that amplitude of
the peaks and there positional relationship are periodic.

Obtained constants $\gamma$ and $\kappa$ demonstrate the universal
nature, they are an attributes of the critical GSK point. For
example, in paper~\cite{Isaeva4} with the help of constant
$\kappa$ the self-similar hierarchical structure of the
complexified H\'{e}non map at the GSK point is demonstrated.

Furthermore, the universal properties of the spectrum at the GSK
point are demonstrated in mentioned physical experiment. With the
aid of the described in~\cite{Isaeva3} device, several first
period-tripling bifurcations  have been successively observed in
physical experiment (to get immediately the GSK point is
impossible problem because of noise and experimental error).
Figure~2~(c) demonstrates the Fourier spectrum of the signal
generated by experimental device with the values of driving
parameters, corresponding to the existence of the cycle of period
$9$ (following period tripled cycles are difficultly observable in
experimental conditions). One can see the high peaks at the
corresponded frequencies. The realization of typical difference
between the amplitudes of intensities of harmonics at the tripling
frequencies in order to constant $\gamma=21.9$ dB is confirmed.

{\it Author is grateful to S.P.~Kuznetsov, and also to
V.I.~Ponomarenko, working up the experimental device. The work is
supported by RFBR, grant No.03-02-16074}

\newpage

\begin{figure}
\centerline{\epsfig{file=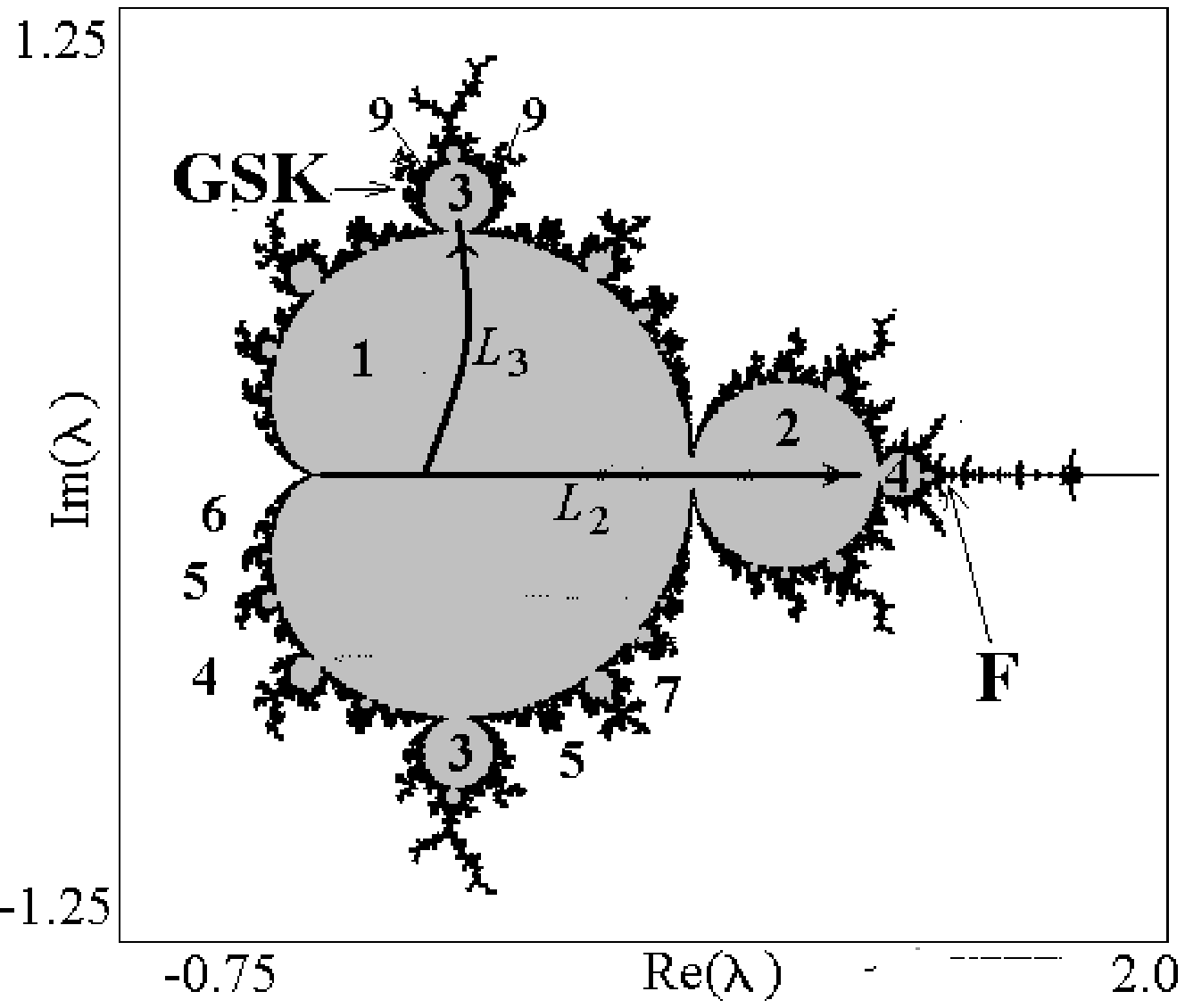,width=0.55\textwidth}}

\caption{Mandelbrot set of the complex map~(1). Gray color
designates the regions of periodic dynamics (periods of cycles are
marked by respective numbers). Black pattern corresponds to the
restricted in the phase space chaotic dynamics. White color means
escape of the orbits to infinity. The line $L_2$ and $L_3$ show
the pathes, following by which one can obtain period-doubling or
period-tripling bifurcation cascades, respectively.}
\end{figure}
\begin{figure}
\centerline{\epsfig{file=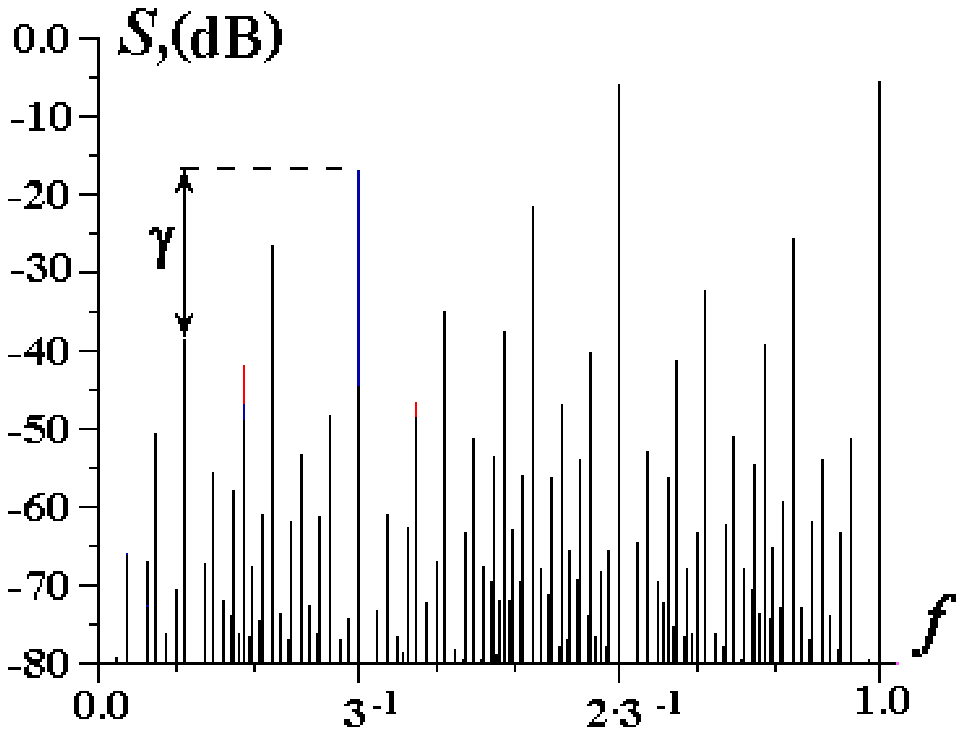,bb=0 0 287 213,width=5.0cm}
\quad \epsfig{file=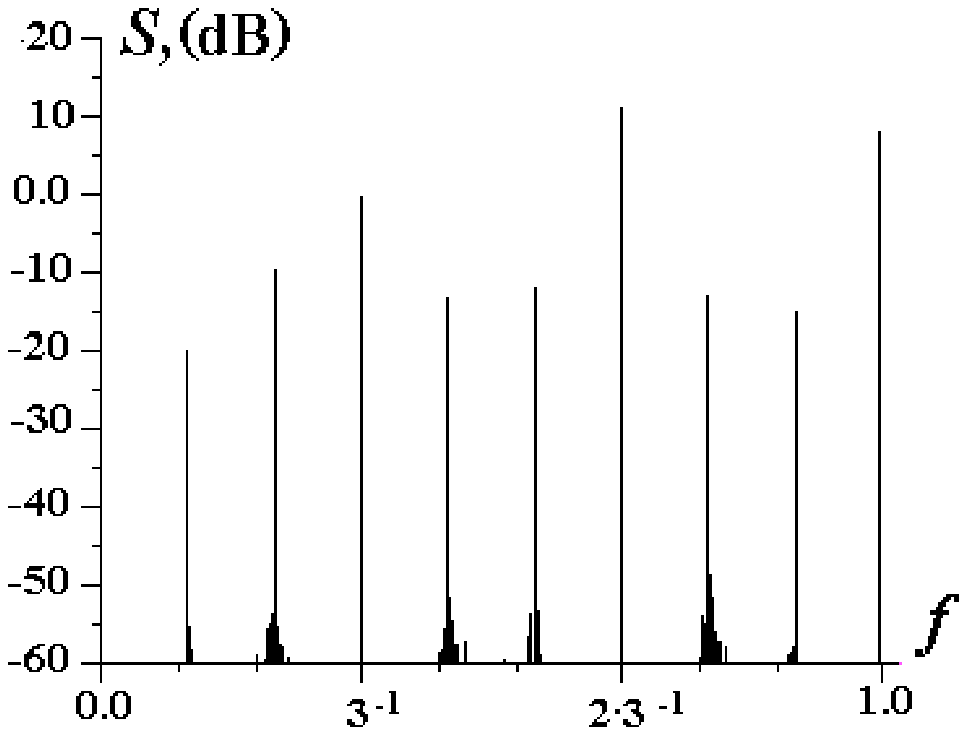,bb=0 0 292 214,width=5.0cm}}

\centerline{a\hspace{4cm}b}

\centerline{\epsfig{file=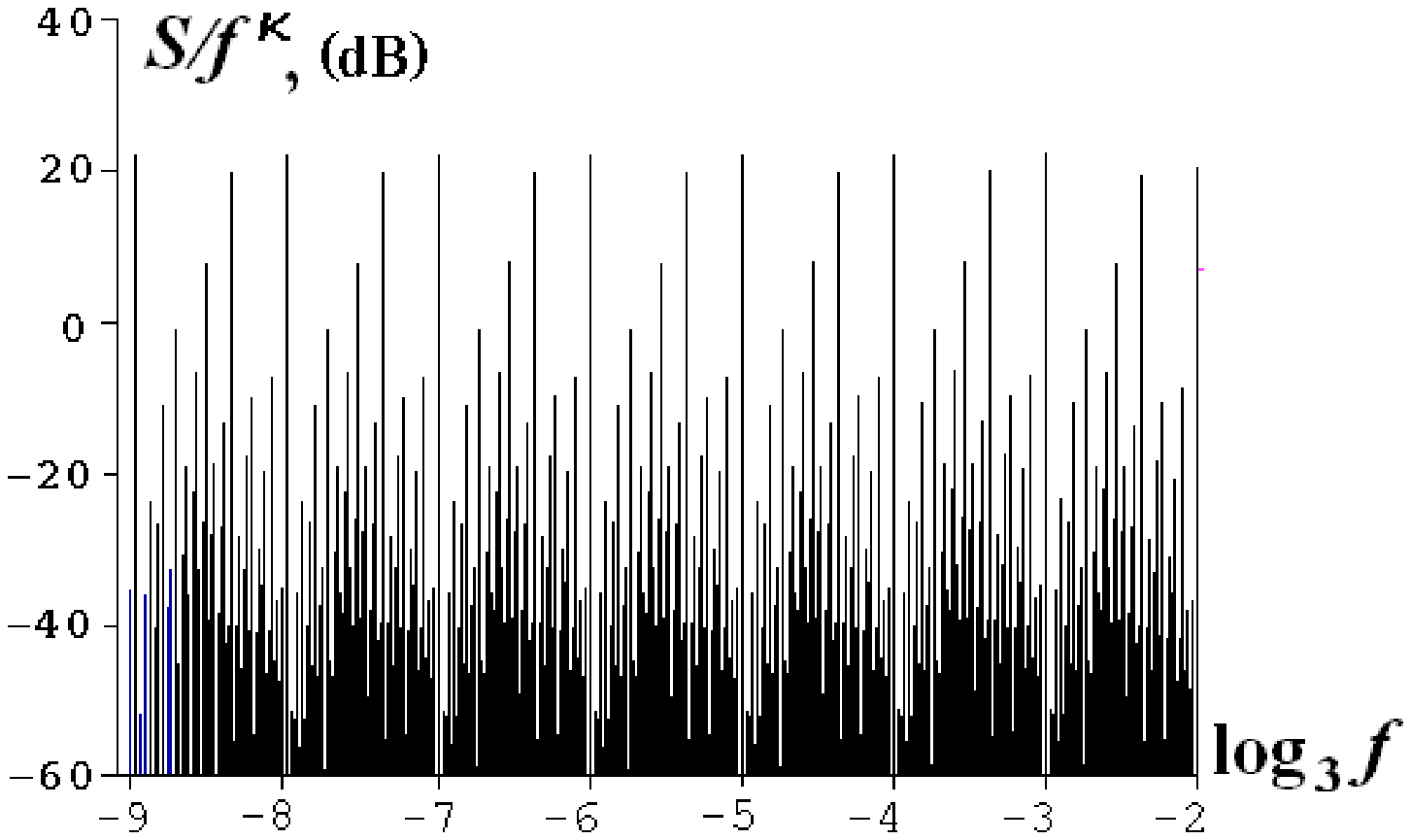,bb=0 0 421 255,width=10.5cm}}

\centerline{c}

\caption{Fourier spectra:~(a) -- spectrum of signal of the complex
map~(1) at the critical point~(2);~(b) -- the same spectrum in
double-logarithmic scale;~(c) -- spectrum of the experimental
signal with period $9$.}
\end{figure}


\newpage

\begin {thebibliography}{9}
\bibitem{Kuznetsov} S.P. Kuznetsov. Dynamical chaos.
M.: Nauka. 2001 (in Russia).

\bibitem{Peitgen} H.-O. Peitgen, P.H. Richter. The
beauty of fractals. Images of complex dynamical
systems. Springer-Verlag, New-York, 1986.

\bibitem{Cvitanovic} P. Cvitanovic, J. Myrheim. Commun. Math. Phys., {\bf121}, No.2,
225-254 (1989).

\bibitem{Golberg} A.I. Golberg, Ya.G. Sinai, K.M. Khanin.
UMN (Adv. Math. Sci.), {\bf38}, No. 1, 159-160 (1983) (in Russia).

\bibitem{Isaeva1} O.B. Isaeva, S.P. Kuznetsov. RCD {\bf5},
No. 4, 459-476 (2000).

\bibitem{Peckham} B.B. Peckham. Int. J. of
Bifurcation and Chaos, {\bf8}, No 1, 73-93 (1998).

\bibitem{Beck} C. Beck. Physica {\bf D125}, 171-182 (1999).

\bibitem{Isaeva2} O.B. Isaeva. Izv. VUZov PND (Appl. Nonlin. Dyn.). V.9, No.6,
129-146 (2001) (in Russia).

\bibitem{Isaeva3} O.B. Isaeva, S.P. Kuznetsov, V.I. Ponomarenko.
PRE. {\bf64}, 055201(4) (2001).

\bibitem{Isaeva4} O.B. Isaeva, S.P. Kuznetsov. Electronic preprint at
http:\\arxiv.org (2005)
\end{thebibliography}
\end{document}